\documentclass[letter,seceq,twocolumn]{jpsj2_122b} 
\def\rnum#1{\expandafter{%
\romannumeral #1}}
\def\Rnum#1{\uppercase\expandafter{%
\romannumeral #1}}

\newcommand{\bol}[1]{\boldsymbol #1}

\title{
Quantum Fluctuations of Chirality in One-Dimensional 
Spin-$\frac{\bol 1}{\bol 2}$ Multiferroics: \\
Gapless Dielectric Response from Phasons and Chiral Solitons
}

\author{
Shunsuke Furukawa$^1$, Masahiro Sato$^1$, Yasuhiro Saiga$^2$, and Shigeki Onoda$^1$
}
\inst{
$^1$Condensed Matter Theory Laboratory, RIKEN, Wako, Saitama 351-0198, Japan\\
$^2$Center for Educational Assistance, Shibaura Institute of Technology, Minuma-ku, Saitama 337-8570, Japan
}

\abst{
We present a quantum theory for one-dimensional spin-$1/2$ multiferroics, 
where the vector spin chirality couples with the electric polarization.  
Based on exact diagonalization and bosonization, it is shown that 
quantum fluctuations appreciably reduce the chiral ordering amplitude 
and the associated ferroelectric polarization. 
This yields nearly collinear spin correlations in short-range scales, 
in qualitative agreement with recent neutron scattering experiments. 
There appear gapless chirality excitations described by phasons and new solitons, 
which can be experimentally verified from the low-energy dielectric response. 
}

\kword{multiferroics, vector spin chirality, quantum spin systems, dielectric response}

\begin{document}
\maketitle

When the chiral symmetry is spontaneously broken by magnetic interactions in Mott insulators, 
a ferroelectric polarization appears through the spin-orbit coupling. 
This new prototype of multiferroic behavior has been discovered 
in a spin-2 helimagnet TbMnO$_3$~\cite{Kimura03}, 
and attracted a current great interest for both its fundamental importance 
and its potential application to an electrical control of spins~\cite{Tokura06,Cheong07}. 
This magnetoelectric coupling 
between the vector spin chirality and the polarization~\cite{KNB1,SergienkoDagotto06,Harris} 
is ubiquitous in Mott insulators~\cite{JONH}. 
Of our particular interest is recently discovered multiferroic behavior 
in one-dimensional (1D) frustrated spin-$1/2$ magnets, 
LiCuVO$_4$~\cite{Enderle05,Naito07,Yasui08,Schrettle08} and LiCu$_2$O$_2$~\cite{Masuda04,Park07,Seki08}.
These have opened an intriguing issue of strong quantum fluctuations in multiferroics 
due to the low dimensionality and the spin-$1/2$ nature,
as suggested by recent neutron scattering experiments~\cite{Seki08,Keimer}.

Another important aspect in multiferroics is 
that low-frequency magnetic excitations can be probed 
from the dielectric functions $\varepsilon^{ii}(\omega)$ through the magnetoelectric coupling
\cite{KNB2,Malashevich08,KOHN}, 
as actually measured for $R$MnO$_3$~\cite{Pimenov06,Kida}. 
Namely, a local flip of the chirality induces the charge dynamics. 
Particularly, in the 1D spin-$1/2$ multiferroics, 
a pair of such local defects may propagate as new elementary excitations. 
This is reminiscent of charged solitons in 1D band insulators, e.g., polyacetylene~\cite{Heeger88}.
However, the chirality-induced charge dynamics in the 1D multiferroics has a much smaller energy scale 
and may produce the low-frequency dielectric response.


In this Letter, we for the first time reveal the novel quantum nature of 1D spin-$1/2$ multiferroics, 
by examining a simple but realistic frustrated spin model.
It is shown that the chiral phase can appear near the $SU(2)$-symmetric case, 
which explains the experimentally observed ferroelectricity.
Significantly, this multiferroic state is characterized by 
(\rnum{1}) a tiny amplitude of the the chiral long-range order (LRO),
(\rnum{2}) almost collinear spin correlations, and 
(\rnum{3}) gapless chiral solitons in the low-frequency dielectric response.

\begin{figure}[b]
  \begin{center}
    \includegraphics[width=8.5cm]{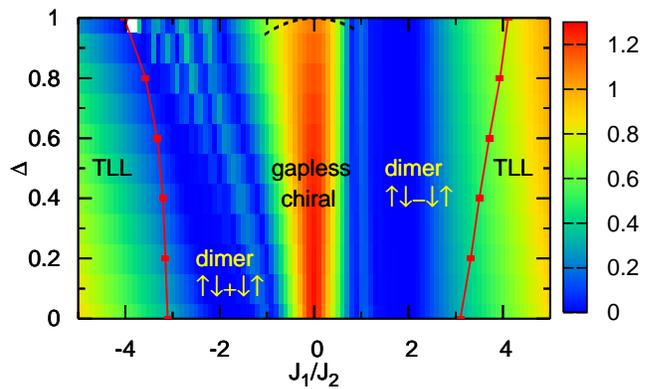}
  \end{center}
\caption{(Color online) 
The intensity plot of the spin Drude weight $D_s/J_2$ for the model~(\ref{eq:H}) with $L=28$. 
The black broken line at $\Delta\approx 1$ shows 
a chiral phase boundary obtained in the scaling analysis. 
The red lines around $|J_1|/J_2 \! \approx \! 3\text{~-~}4$ are the phase boundaries between TLLs and dimer phases 
determined in refs.~\citen{Nomura94} and \citen{Somma01}.
At $J_1=0$, there is no chiral order.
A dark region with $J_1<0$ contains at least two phases:
for small $\Delta$, a dimer order appears~\cite{Somma01} with a unit 
$|\!\!\uparrow\downarrow\rangle+|\!\!\downarrow\uparrow\rangle$ 
on the nearest-neighbor bonds, while for $\Delta\approx 1$,
another gapped phase is expected to appear~\cite{Itoi01}.
}
\label{fig:phase_XXZ}
\end{figure}

The minimal model for the 1D spin-$1/2$ multiferroics, especially for LiCuVO$_4$~\cite{J1J2}, 
is given by a simple spin Hamiltonian 
with nearest- and second-nearest-neighbour exchange couplings, $J_1$ and $J_2$:
\begin{equation}
  H={\sum_{n=1}^2\sum_j} J_n
   \left[{\bol S}_j\cdot{\bol S}_{j+n}+(\Delta-1)S^z_jS^z_{j+n}\right].
\label{eq:H}
\end{equation}
Here, ${\bol S}_j$ represents the spin operator on the site $j~(=1,\dots,L)$ 
and $\Delta(\le1$) is an easy-plane exchange anisotropy. 
In the classical limit, 
a chiral ground state is realized when $|J_1|/J_2<4$, 
having a nonzero expectation value $\kappa^z=\langle {\cal O}_{\kappa}^z\rangle$ 
of the vector spin chirality 
\begin{equation}
  {\cal O}_{\kappa}^\alpha\equiv L^{-1}{\sum_{j}} ({\bol S}_j\times{\bol S}_{j+1})^\alpha,~~
  \alpha=x,y,z. 
\label{eq:O_k}
\end{equation}

In the quantum case with spin-$1/2$, 
the existence of a chiral phase with an incommensurate gapless excitation 
has been proposed based on the bosonization combined with a mean field theory~\cite{Nersesyan98}, 
and has been confirmed numerically around the XY case $\Delta\approx 0$~\cite{Hikihara01}.
On the other hand, in the $SU(2)$-symmetric case $\Delta=1$, 
the chiral ordering is prohibited, and 
valence bond solids (VBS) are stabilized~\cite{White96,Itoi01}. 
The intermediate regime of the anisotropy $0<\Delta<1$ has not been clarified yet.
%
We revisit this issue by calculating the spin Drude weight $D_s$ by Lanczos diagonalization~\cite{Drude}, 
to see whether the spin excitation is gapless or not; see Fig.~\ref{fig:phase_XXZ}.
We find a region with moderate $D_s/J_2$ for $|J_1|/J_2\lesssim 1$, 
which can be assigned to the gapless chiral phase found in refs. \citen{Nersesyan98} and \citen{Hikihara01}.
When increasing $|J_1|/J_2$ from zero in the XY case $\Delta=0$, 
$D_s/J_2$ starts to decrease rapidly at $|J_1|/J_2 \approx 0.9$, 
in reasonable agreement with the previous numerical results on the phase boundary 
between the chiral and dimer phases~\cite{Hikihara01,Okunishi08}. 
When increasing $\Delta$ from zero for $|J_1|/J_2\lesssim 1$, $D_s/J_2$ decreases only gradually, 
which suggests that the gapless chiral phase might extend to the vicinity of $\Delta=1$.
On the other hand, it is known\cite{Itoi01} that 
the gap associated with the VBS states for $\Delta=1$ and $|J_1|/J_2\lesssim 1$ 
is too small to detect in numerical analyses of small clusters.



To correctly determine the phase diagram for small $|J_1|/J_2$, 
we present a scaling analysis based on the bosonization.
We start from the bosonized effective Hamiltonian of eq.~\eqref{eq:H} 
formulated in ref.~\citen{Nersesyan98} for $|J_1|/J_2\ll 1$, 
which consists of two Tomonaga-Luttinger liquid (TLL) parts ${\cal H}_\pm[\Phi_\pm,\Theta_\pm]$ 
and interaction terms ${\cal H}_{\rm int}$.
Here $(\Phi_\nu,\Theta_\nu)$ is the pair of dual boson fields. 
The TLL parameters of the sectors ${\cal H}_\pm$ are obtained as 
\begin{eqnarray}
 K_\pm \approx 
 K\left[ 1\mp K\frac{J_1\Delta a_0}{2\pi v}+O\left( \left(\frac{J_1}{J_2}\right)^2 \right) \right], 
\label{eq:TLLpara}
\end{eqnarray}
where $K$ and $v$ are the TLL parameter and the spin-wave velocity for $J_1=0$, respectively.
With the easy-plane anisotropy $\Delta<1$, 
two interaction terms 
${\cal O}_c=(\partial \Theta_+)\sin(\sqrt{4\pi}\Theta_-)$ with the scaling dimension $\Delta_c=1+1/K_-$ 
and 
${\cal O}_d=\cos(\sqrt{4\pi}\Phi_+)\cos(\sqrt{4\pi}\Theta_-)$ with the dimension $\Delta_d=K_+ + 1/K_-$ 
can be relevant for $J_1>0$, 
while only ${\cal O}_c$ can for $J_1<0$. 
Here, a relevant ${\cal O}_{c(d)}$ can yield a gapless chiral (singlet-dimer) order. 
Therefore, we obtain the following criterion: 
for $J_1>0$, the phase boundary between dimer and chiral phases is fixed by the relation $\Delta_c=\Delta_d$,
while for $J_1<0$, the chiral order appears under the condition $\Delta_c<2$.
Then the chiral phase boundary is given by 
\begin{equation}
\Delta\approx 1-(J_1/J_2)^2/(2\pi^2)+\cdots
\end{equation}
for $|J_1|/J_2 \ll 1$,
as indicated by the black broken line in Fig. \ref{fig:phase_XXZ}. 
Namely, a small easy-plane anisotropy, 
which is realistic for spin-$1/2$ systems, 
induces the chiral order. 
We also calculate the one-loop RG corrections to $K_{\pm}$,
which suggest that the chiral phase is robust for $J_1<0$ 
while it is reduced by an expansion of the dimer phase for $J_1>0$.  
This tendency is in accord with the slower decay of $D_s/J_2$ on the ferromagnetic side $J_1<0$ 
in Fig.~\ref{fig:phase_XXZ}.
Henceforth, we focus on the realistic case $\Delta\approx 1$ for the cuprates.

Now let us investigate properties of the chiral ground state, 
i.e., quantum multiferroics. 
The uniform chirality $\kappa^z$ produces the electric polarization $P^y$
through the inverse Dzyaloshinskii-Moriya (DM) 
interaction~\cite{KNB1,SergienkoDagotto06,Harris,JONH}, i.e., 
$P^\alpha\propto \epsilon_{\alpha\beta x} {\cal O}_\kappa^\beta$, 
where $\epsilon_{\alpha\beta\gamma}$ is the fully antisymmetric tensor.
Then it also generates the lattice modulation 
due to the electrostatic force. 
We take into account this coupling between spins and transverse optical phonons 
through the DM interaction~\cite{OnodaNagaosa07}. 
Integrating out phonons gives a biquadratic DM interaction,
\begin{equation}
H_\mathrm{BDM}=-V_z
{ \sum_j}
({\bol S}_j\times{\bol S}_{j+1})^z
({\bol S}_{j+2}\times{\bol S}_{j+3})^z.
\label{eq:H_V}
\end{equation}
Here, longer-range terms have been neglected for simplicity.
This term yields a gapless chiral phase for quantum spin ladders~\cite{Sato07} 
and a spin cholesterics for classical spin systems~\cite{Villain77,OnodaNagaosa07}.


\begin{figure}[b]
\begin{center}
\includegraphics[width=8.5cm]{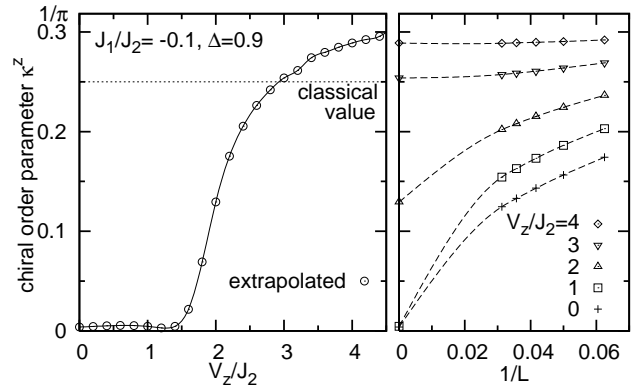}
\end{center}
\caption{
The left panel shows 
the chiral order parameter 
$\kappa^z=\sqrt{\langle ({\cal O}_{\kappa}^z)^2 \rangle}$ 
for $J_1/J_2=-0.1$ and $\Delta=0.9$ as a function of $V_z/J_2$. 
Extrapolation is done from the sequence of data for $L=16,20,24,28$ and $32$, 
using the alternating $\epsilon$-algorithm \cite{Barber83}. 
On the right panel, the extrapolated values (placed at ``$1/L=0$'') 
are plotted together with the finite-size data. 
Broken lines are guides to eyes.
}
\label{fig:ChiralOrder}
\end{figure}

\newcommand{\Scal}{{\mathcal S}}
\newcommand{\Ccal}{{\mathcal C}}
First, the chiral order parameter calculated from 
$\kappa^z=\sqrt{\langle ({\cal O}_{\kappa}^z)^2 \rangle}$ 
is shown in Fig.~\ref{fig:ChiralOrder}.
The finite-size results are extrapolated to $L\to\infty$ 
by the alternating $\epsilon$-algorithm \cite{Barber83}.
We take a small $|J_1|/J_2$ 
to assure that the system is inside the chiral phase even at $V_z=0$ 
and to make the incommensurate wave vector $Q$ sufficiently close to $\pi/2$.
Remarkably, when $V_z/J_2 \lesssim 2$ for $\Delta=0.9$,
$\kappa^z$ is significantly suppressed from the classical value $S^2=1/4$, 
indicating appreciable quantum fluctuations of the chirality near the $SU(2)$-symmetric case.
This is consistent with the bosonization analysis combined with the mean-field theory~\cite{Nersesyan98},
which gives $\kappa^z\sim |J_1/J_2|^{1/(K_--1)}$ with $K_--1\sim \sqrt{1-\Delta}$ near $\Delta=1$. 
On the other hand, with increasing $V_z$,
$\kappa^z$ starts to increase rapidly around $V_z/J_2 \sim 2$ and approaches a constant $1/\pi$, 
which is the maximal value for the spin-$1/2$ case.


\begin{figure}[b]
\begin{center}
\includegraphics[width=8.5cm]{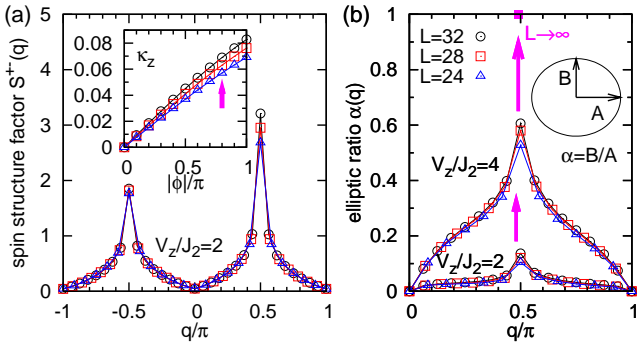}
\end{center}
\caption{(Color online)
(a) The spin structure factor $\Scal^{+-}(q)$ 
for $J_1/J_2=-0.1$, $\Delta=0.9$, and $V_z/J_2=2$ with a flux $\phi=-0.8\pi$.
The insets show the $\phi$-dependence of $\kappa^z=\langle{\cal O}_\kappa^z\rangle$.
(b) The elliptic ratio $\alpha(q)$ for each $q$ component of the spiral correlations, 
for the same $J_1/J_2$ and $\Delta$ and for $V_z/J_2=2$ and $4$.
}
\label{fig:StructureFactor}
\end{figure}

The weak chirality for $V_z/J_2\lesssim 2$ has a remarkable consequence 
on the spin correlations.
Figure~\ref{fig:StructureFactor} (a) shows
the equal-time in-plane spin correlation,
$
 \Scal^{+-}(q)=\langle S^+_q S^-_{-q} \rangle ~~
 \text{with}~S^\alpha_q= {\sum_j} S^\alpha_j e^{-iqj}/\sqrt{L}
$
for $V_z/J_2=2$.
To break the inversion symmetry in finite-size systems 
and to have a chiral order parameter comparable to the value 
extrapolated to $L\to\infty$ in Fig.~\ref{fig:ChiralOrder}, 
a flux $\phi=-0.8\pi$ 
is inserted through the periodic chain as a mean field.
Figure~\ref{fig:StructureFactor} (a) shows two comparable peaks at $q=\pm Q$ 
and a tiny asymmetry under the inversion $q\to -q$, reflecting a small chirality.
According to the bosonization analysis of the gapless chiral phase \cite{Nersesyan98}, 
only the peak at $q=Q$ eventually diverges in the thermodynamic limit, due to a spiral quasi-LRO.
The other peak at $q=-Q$ observed for $V_z/J_2=2$ converges to a large but finite value. 
Nevertheless, except in the close vicinity of $q=\pm Q$, 
the asymmetry is small for $V_z/J_2\lesssim 2$.
%
To understand the physical meaning of these results, 
it is useful to introduce 
\begin{equation}
 \alpha(q)\equiv \frac{\sqrt{\Scal^{+-}(q)}-\sqrt{\Scal^{+-}(-q)}}
                 {\sqrt{\Scal^{+-}(q)}+\sqrt{\Scal^{+-}(-q)}}
\end{equation}
for each $q$ component, 
which represents the ratio of the minor to major axes of ellipse.
When $\alpha=0$ ($\alpha=1$), the spin correlation is collinear (a circular spiral). 
The divergence of ${\cal S}^{+-}(Q)$ indicates $\alpha (Q)=1$ in the thermodynamic limit. 
On the other hand, away from $q=Q\approx\pi/2$, 
the elliptic ratio $\alpha (q)$ converges to a rather small value,
particularly for smaller $V_z/J_2$, as shown in  Fig.~\ref{fig:StructureFactor} (b).
Therefore, the spin correlations are elliptic, 
even nearly collinear, at short-range scales 
and only the quasi-LRO component is characterized by a circular spiral with a pitch $Q$.
The circular to collinear crossover in the length scale also appears in the time scale.
Namely, the dynamical structure factor $\Scal^{+-}(Q,\omega)$ diverges algebraically 
as $\sim 1/\omega^{2-1/(2K_+)}$
and our bosonization analysis shows that 
an energy gap $\varepsilon_g\sim J_2|J_1/J_2|^{K_-/(K_--1)}$ opens at $q=-Q$, 
which is tiny near the $SU(2)$-symmetric case.
Thus, ${\cal S}^{\pm\mp} (q,\omega)$ shows a {\em circular} spiral
only for $\omega, v||q|-Q|\lesssim\varepsilon_g$, 
and is nearly {\em collinear} otherwise.

This provides a key to understanding puzzling results 
from a recent elastic polarized-neutron scattering experiment in LiCu$_2$O$_2$~\cite{Seki08}:
the elastic elliptic ratio $\approx 9\text{~-~}20\%$
indicating an {\em elliptic} spiral, and
$\Scal^{xx}(Q, 0)\approx \Scal^{yy}(Q,0)$ indicating a {\em circular} spiral,
if one interprets that the observed intensity originates solely from the genuine Bragg peak 
due to a 3D magnetic LRO.
This problem can be resolved, at least qualitatively, 
if the fine but finite energy-momentum resolution 
allows an additional contribution from low-energy spin fluctuations to the elastic intensity
(see also a related argument in ref. \citen{KOHN}).
Actually, according to our analysis of the weak-chiral region $V_z/J_2\lesssim 2$, 
there exist large spin fluctuations outside the narrow window $\omega, v||q|-Q|\lesssim\varepsilon_g$,
which show small $\alpha$ but retain $\Scal^{xx}=\Scal^{yy}$ even at high $\omega$.
Detailed inelastic polarized neutron scattering experiments are required
to test this possibility and to reveal nontrivial effects of a weak 3D coupling on the spin dynamics.

\begin{figure}[b]
\begin{center}
\includegraphics[width=8.5cm]{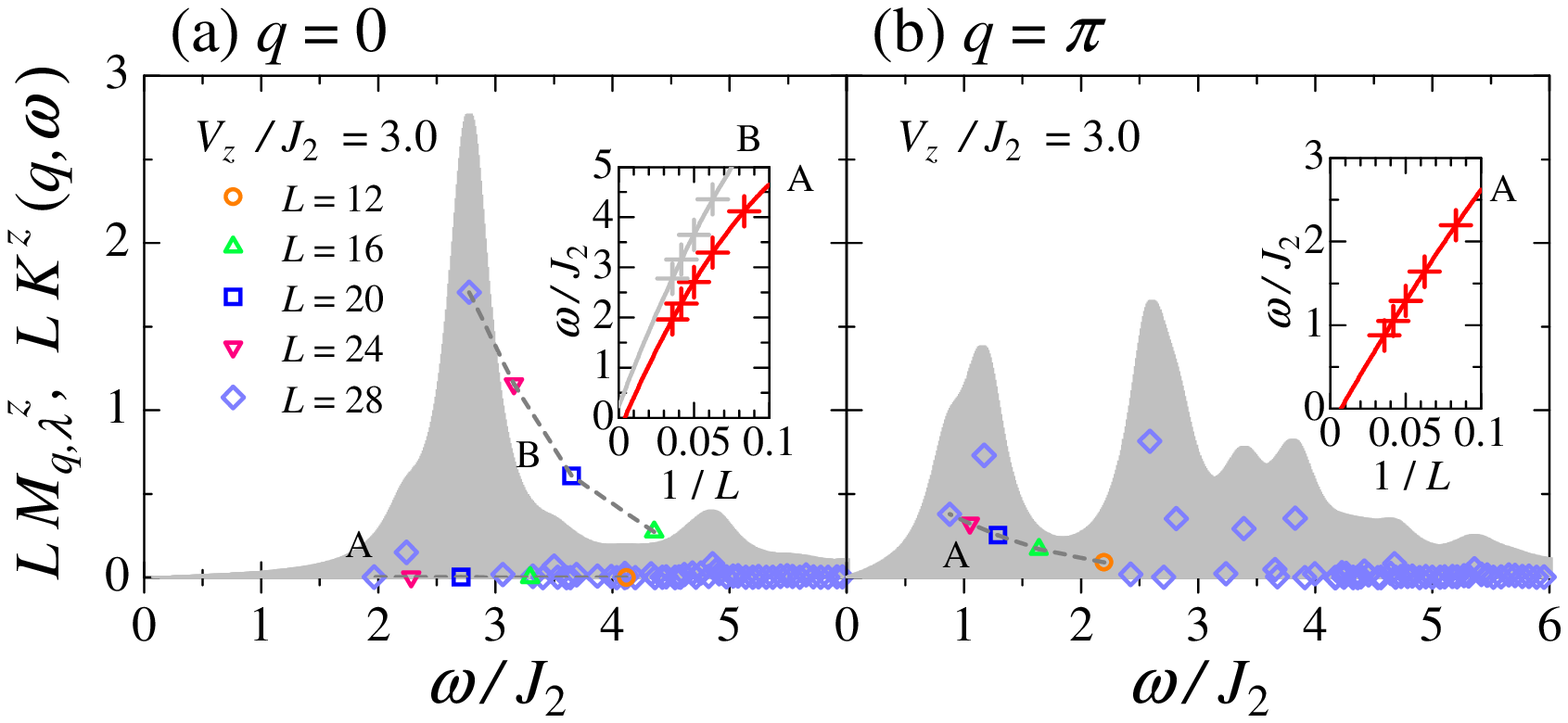}
\includegraphics[width=8.5cm]{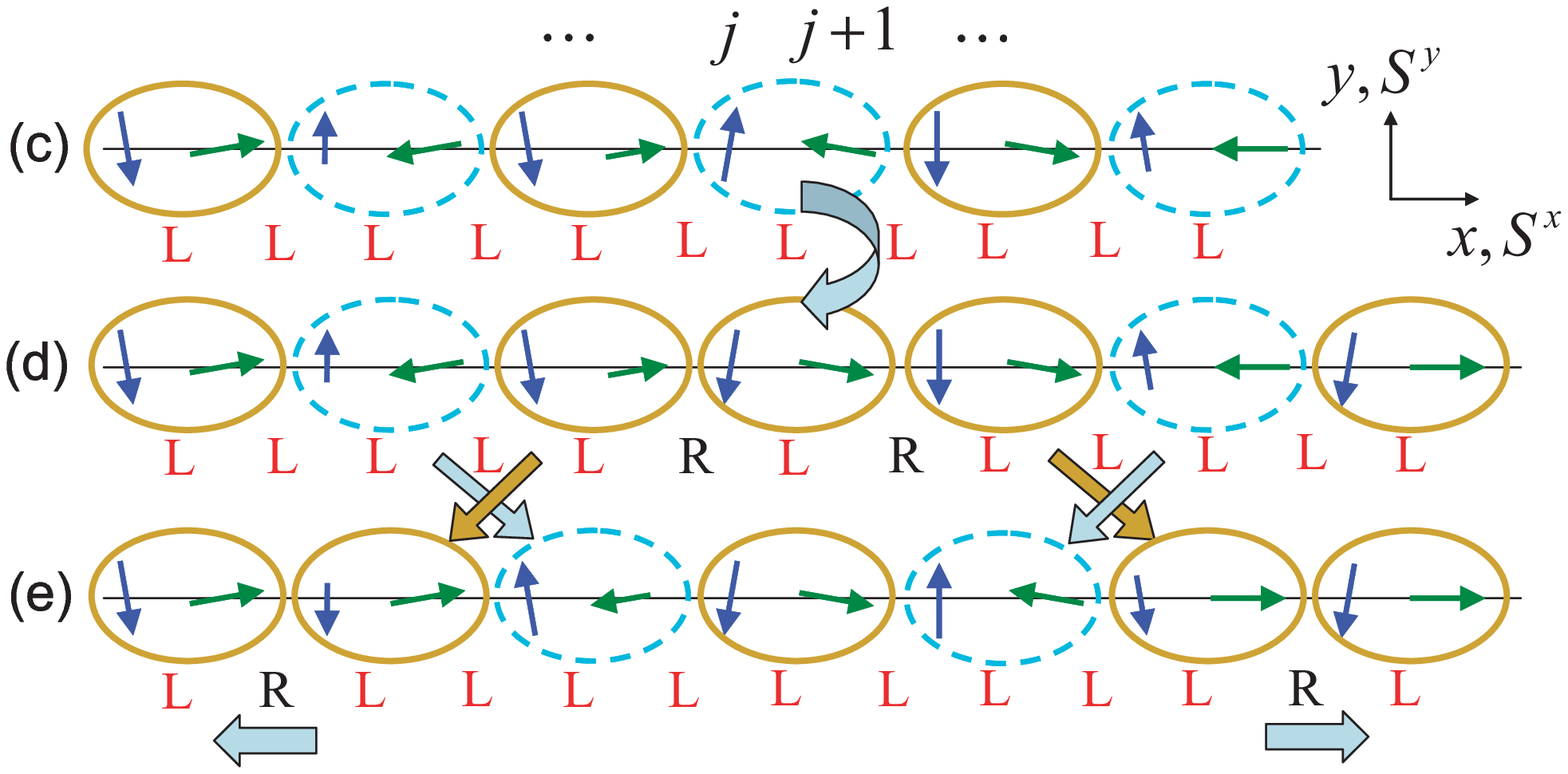}
\end{center}
\caption{(Color online) 
 Dynamical in-plane chirality correlation function $K^z(q, \omega)$
 at (a) $q=0$ and (b) $q=\pi$, respectively, 
 for $J_1/J_2=-0.1$, $\Delta=0.9$, and $V_z/J_2=3$. 
 The shaded areas show the spectra for $L=28$ 
 drawn with $\eta = 0.2 J_2$ (see eq.~\eqref{eq:ch_dy}). 
 In (a), the pseudo ground state is removed since it does not contribute to the dynamics in the thermodynamic limit.
 The symbols show the matrix elements
 $
   L M_{q,\lambda}^z 
     = L^2 | \langle \lambda | 
              {\cal O}_\kappa^z (q)^\dagger 
           | {\rm GS} \rangle |^2
 $ 
 at $\omega = \omega_\lambda$, where $| \lambda \rangle$ denotes
 an eigenstate with energy $\omega_\lambda$ measured from the (true) ground state.
 For $L=28$, they are shown for all $\lambda$'s; 
 for smaller $L$, they are shown only for the low-energy levels depicted as $A$ and $B$.
 The insets show the extrapolations of the $A$ and $B$ levels into $L\to\infty$. 
 (c-e) Schematic pictures on the creation and propagation of the chiral soliton pair 
 in the background of quasi-long-range spiral order. 
 The symbol $L$ ($R$) denotes a positive or {\em left-handed} 
(negative or {\em right-handed}) chirality at each bond.
}
\label{fig:chiraldynamics}
\end{figure}

Next, we discuss dynamical in-plane and out-of-plane chiral correlations, 
$K^z$ and $K^y$, respectively;
\begin{equation}
\label{eq:ch_dy}
K^\alpha (q,\omega) = 
-\frac L\pi 
~\Im 
  \langle \text{GS}| {\cal O}_\kappa^\alpha(q) 
 \frac{1}{\omega-{\cal H}+i\eta} 
 {\cal O}_\kappa^\alpha(q)^\dag |\text{GS}\rangle,
\end{equation}
where $\eta\to +0$ and 
$
 {\cal O}_\kappa^\alpha(q) 
 = L^{-1}\sum_j e^{-iqj}({\bol S}_j\times{\bol S}_{j+1})^\alpha
$ 
is the Fourier component of the chirality. 
There appear the following contributions to dielectric functions $\varepsilon^{\alpha\beta}(q,\omega)$ 
through the magnetoelectric coupling: 
$\Im\varepsilon^{yy} (q,\omega) \propto K^z (q,\omega)$ and 
$\Im\varepsilon^{zz} (q,\omega) \propto K^y (q,\omega)$.
%
We have performed numerical continued-fraction analyses of Lanczos results 
on $K^\alpha(q,\omega)$ at $q=0,\pi$. 
Then, it is found that $K^y(0,\omega)$, 
which was argued in the context of electromagnons~\cite{KNB2}, 
and $K^y(\pi,\omega)$ are gapful. 
On the contrary, 
both $K^z(0,\omega)$ and $K^z(\pi,\omega)$ are gapless, as we discuss below.
To be explicit, we show the numerical results on $K^z(0,\omega)$ and $K^z(\pi,\omega)$
in Figs.~\ref{fig:chiraldynamics} (a) and (b), respectively, 
for the same $J_1/J_2$ and $\Delta$ 
as used in Fig.~\ref{fig:ChiralOrder} but for $V_z/J_2 = 3$.
In the insets, we clarify finite-size effects on the energy levels 
of the first excited states ($A$) and a higher-energy state ($B$) with large intensity.
They indicate that both chirality excitations $K^z(0,\omega)$ and $K^z(\pi,\omega)$ 
eventually become gapless in the thermodynamic limit, 
leading to a gapless dielectric response even in a Mott insulator.
This indicates an illuminating feature that 
this 1D multiferroic state possesses an antiferro-chiral (and thus antiferroelectric) quasi-LRO, 
in addition to the uniform chiral LRO.


Now we clarify the origin of the above gapless chirality dynamics
using the bosonized expressions of the chirality operators in the gapless chiral phase:
\begin{subequations}
 \label{boson_formula}
 \begin{align}
  \label{boson_formula_1}
  {\cal O}_\kappa^z(0) &\sim \int\!\! \frac{dx}{La_0} 
   \left[ c_1\sin(\sqrt{4\pi}\Theta_-) + c_2 a_0^2\kappa^z (\partial_x\Theta_+)^2+\dots\right], \,\,\,\,\,\\
  \label{boson_formula_2}
  {\cal O}_\kappa^z(\pi) &\sim \int\!\! 
  \frac{dx}{La_0} ~c_3\kappa^z\cos(\sqrt{4\pi}\Phi_+)+\cdots,\,\,\,\,\, 
 \end{align}
\end{subequations} 
where $c_{1-3}$ are $O(1)$ constants and $a_0$ is a lattice spacing.
%
%
%
Note that excitations associated with $(\Phi_-,\Theta_-)$ are gapful 
because of the chiral LRO, 
while those associated with $(\Phi_+,\Theta_+)$ are gapless 
since the spin rotational symmetry is conserved. 
Let us begin with the uniform part. 
The first term in eq.~\eqref{boson_formula_1} gives rise to gapful spinon-pair excitation in $K^z(0,\omega)$, 
which is assigned to the shaded spectrum 
above the energy level $B$ in Fig.~\ref{fig:chiraldynamics}~(a).
On the other hand, the second term in eq.~\eqref{boson_formula_1} 
gives a nonsingular gapless contributions to $K^z(0,\omega)$, 
which explains the spectrum from $A$ in Fig.~\ref{fig:chiraldynamics}~(a). 
This gapless mode turns to two-phason excitations 
if a 3D incommensurate spiral magnetic LRO appears due to a weak inter-chain coupling.  
In realistic systems containing defects that act as pinning centers, 
single phason excitation can also appear in $K^z(0,\omega)$.\cite{Fukuyama78}


Similarly, the term $\cos(\sqrt{4\pi}\Phi_+)$ in the staggered part ${\cal O}_\kappa^{z}(\pi)$ 
also yields a gapless mode, 
which leads to power-law behavior $K^z(\pi,\omega)\sim \omega^{2K_+-2}$; 
see Fig~\ref{fig:chiraldynamics}~(b). 
The physical origin is understood by noticing the following properties: (\rnum{1}) 
$\cos(\sqrt{4\pi}\Phi_+)$ translates $\Theta_+$ in $S^+_j\propto e^{i\sqrt{\pi}\Theta_+}$ by $\sqrt{\pi}$, and 
(\rnum{2}) $\int\!dx\cos(\sqrt{4\pi}\Phi_+)$ coincides with the 
leading term in the bosonized form of $\sum_j S^z_jS^z_{j+1}$, 
which rotates the two neighbouring spins by $\pi$ about the $z$ axis.
Accordingly, the gapless excitations in $K^z(\pi,\omega)$
consist of two solitons, as schematically shown in
Fig.~\ref{fig:chiraldynamics} (c)-(e): 
An operation of $\cos(\sqrt{4\pi}\Phi_+)$ transforms a state from (c) to (d). 
It creates a pair of kinks, which contain, 
for instance, a negative ($R$) chirality in the positive ($L$) chiral background.
Hence, they induce the opposite transverse displacement of electric charge
in comparison with the ferroelectric background.
Then, via the exchange of two neighboring chiral spin pairs denoted by circles, 
the kinks propagate without changing the sign of the chirality of each domain, as shown in (e).
This soliton results from a chiral pairing of two spinons,
in contrast to the gapless charge-neutral spinon in 1D spin-$1/2$ antiferromagnets. 
The transverse nature of the charge displacement also contrasts with 
the longitudinal nature of the well-known charge soliton in the polyacetylene~\cite{Heeger88}.
In the 1D multiferroics like LiCuVO$_4$ 
where the unit cell is doubled, 
the chiral solitons observed in the staggered part 
can also appear in the uniform part. 
If the $U(1)$ symmetry is weakly broken 
by spin anisotropy or an in-plane magnetic field, 
the soliton can acquire a small gap. 
It will be interesting to look for this soliton-pair excitation 
by low-frequency electrical or optical probes.

In real materials, an inter-chain coupling $J^\prime$ causes the 3D spiral LRO. 
However, a low transition temperature $T_N \approx 2K$ in
LiCuVO$_4$~\cite{Naito07,Yasui08,Schrettle08,Enderle05} 
suggests that the effect of $J^\prime$ is small and 
should not alter our main conclusions, particularly on the quantum dynamics.

The authors thank A. Furusaki, H. Katsura, N. Kida, 
T. Momoi, N. Nagaosa, S. Seki, Y. Tokura 
for stimulating discussions. 
The work was partly supported by Grants-in-Aid for Scientific Research 
(No. 17071011, No. 20029006, and No. 20046016) 
from MEXT of Japan.
The calculations were partly performed using supercomputers 
in ISSP, University of Tokyo.


\newcommand{\etal}{{\it et al.}}
\newcommand{\PRL}[3]{Phys. Rev. Lett. {\bf #1} (#3) #2}
\newcommand{\PRLp}[3]{Phys. Rev. Lett. {\bf #1} (#3) #2}
\newcommand{\PRB}[3]{Phys. Rev. B {\bf #1} (#3) #2}
\newcommand{\PRBp}[3]{Phys. Rev. B {\bf #1} (#3) #2}
\newcommand{\PRBR}[3]{Phys. Rev. B {\bf #1} (#3) #2 (R)}
\newcommand{\arXiv}[1]{arXiv:#1}
\newcommand{\JPSJ}[3]{J. Phys. Soc. Jpn. {\bf #1} (#3) #2}
\newcommand{\PTPS}[3]{Prog. Theor. Phys. Suppl. {\bf #1} (#3) #2}

\end{document}